# A Portal Analysis for the Design of a Collaborative Research Environment for Students and Supervisors (CRESS) within the CSCR Domain

Vita Hinze-Hoare
School of Electronics and Computer Science
University of Southampton
United Kingdom
2007

**Abstract**
In a previous paper the CSCR domain was defined. Here this is taken to the next stage where we consider the design of a particular Collaborative Research Environment to support Students and Supervisors CRESS. Following the CSCR structure a preliminary design for CRESS has been established and a portal framework analysis is undertaken in order to determine the most appropriate set of tools for its implementation.

## 1. Introduction

In a previous paper (Hinze-Hoare 2006) a definition of the CSCR environment was provided which demonstrated that the CSCW and CSCL environment by themselves are not rich enough to encompass the requirements of collaborative research. An additional five research spaces were identified as necessary components for a CSCR environment.

In this paper the application of the CSCR domain to the specific needs of supporting collaborative research students and their supervisors (CRESS) will be analysed with a view to obtaining the specific set of tools required for the design of the CRESS interface. What follows is a summary of a detailed analysis of the advantages and disadvantages of a range of tools drawn from the analysis of a range of e-laboratories and considered in the light of the requirements of the particular CSCR environment to support research students and their supervisors.

13 working environments and three learning environments have been analysed. In order to determine the most relevant tools which might be applied in the construction of the CRESS environment an analysis has been performed (Hinze-Hoare 2007) which was based upon an assessment of advantages and disadvantages for each tool set with reference to the needs of collaborative research. The final set of tools is instanced in Table1 which summarises the toolset to be employed initially in the new CRESS environment.

| CATEGORY SPACES | TOOLS | Required | Not Required | Review |
|---|---|---|---|---|
| Administration Space (including security tools) | Login | X | | |
| | Access/authorisation Tools | X | | |
| | Recording /Replay Facility | X | | |
| | Instant Messaging Recording | | | X |
| | Assistive Agent | | X | |
| | Help Pane | X | | |
| | Information Link Map | | X | |
| | Scenario/Control flow Tools | | X | |
| Communication tools (including Identification tools) | Text/ Chat | X | | |
| | Audio/Voice | X | | |
| | Still Picture | X | | |
| | Video | X | | |
| | Instant Messaging | | | X |
| | Forum | X | | |
| | Message Board/News | X | | |
| | Avatar (Representations) | | X | |
| | Presence Indicator/Information | X | | |
| | Location Identifier | | X | |
| | Focus Indication | X | | |
| | Participant Data | X | | |
| Scheduling tools | Scheduling Tool | X | | |
| | Task Setting | X | | |
| | Task Monitoring | X | | |
| Shared working space | Whiteboard | X | | |
| | Collaborative Working Window | X | | |
| | 3D Environment | | X | |
| Product Space | Output Window | X | | |
| | Simulations | | X | |
| Reflection Space | Reflective Journal/Private | X | | |
| Social Interaction Space | Community Creation | X | | |
| | Tags (marking Content) | X | | |
| | Friend (file sharing) | X | | |
| | Blog (Public + Private) | X | | |
| | RSS feed to centralize data | X | | |
| Assessment / Feedback Space | Assessment | X | | |
| | Feedback | X | | |
| Supervisor/Tutor Space | Private area for tutors | X | | |
| Knowledge Space | Contribution Database | X | | |
| | Academic database | X | | |
| | Depository | X | | |
| | PowerPoint Slides/Notes | X | | |
| Privacy | Private Space | X | | |
| Public | Public information space | X | | |
| Negotiation | Peer Review assistance | X | | |
| Publication | Schemas/Templates | X | | |
| | Publishing assistance | x | | |

**Table 1 Summary of tools required for deployment in the CRESS environment**

The CSCR Domain has been defined in such a way as to enable analysis and design of many specific and individual interfaces to be constructed for a range of collaborative research environments. The analysis of the requirements for the specific interface has been considered in detail namely a collaborative research environment for the support of students and supervisors CRESS.

A specific toolset for CRESS has been arrived at which will initially be incorporated into a storyboard for user analysis. Specific tools for the CRESS environment are expected to be available to some degree in the range of open source portal frameworks.

**Portal Framework Analysis**
There are a wide number of portal frameworks available for the development of the CRESS interface. A brief survey according to cms.matrix.org shows that there are over 500 portal software developments suites. Not all of these are suitable however for a collaborative virtual research environment. A short list of 10 portal frameworks, some independent of the cms matrix, has been produced and analysed according to the criteria which has been laid out for the CSCR environment.

Table 4 (Appendix) shows the analysis portal frameworks against CSCR category tools. It can be seen that the most closely matching portal framework is Sakai with 39 matching points.

A summary of all the matching points is shown in the portal frequency analysis see table 2. It is clear that the Sakai/Agora Framework has almost twice as many matching points as the next nearest Portal framework analysed.

**Gap Analysis: Sakai/Agora**
Although the Sakai Framework has the highest score of 39 points it is nevertheless important to perform a Gap analysis to find out exactly which tools required for CRESS are already available and which would need to be customised. The results of this can be seen in Table3 (Appendix).

This analysis reveals that all tools are already available in the Sakai/Agora Portal Framework except for:
- Tags (marking Content)
- Contribution Database
- Peer Review assistance
- Publishing assistance

These tools have been shown to be essential to the functionality of the CRESS environment and if they cannot be found as ready made portlets they will need to be constructed from scratch for the purpose of completing the full research environment.

| Portal Framework | Matching Points |
|---|---|
| Sakai | 39 |
| Elgg | 20 |
| Oracle Portal | 15 |
| Light Portal | 13 |
| DotNetNuke | 13 |
| Gridsphere | 13 |
| Ugforge | 11 |
| Liferay | 8 |
| JBoss | 6 |
| J Porta | 6 |

**Table 2 Portal Framework Frequency Analysis**

**Summary**

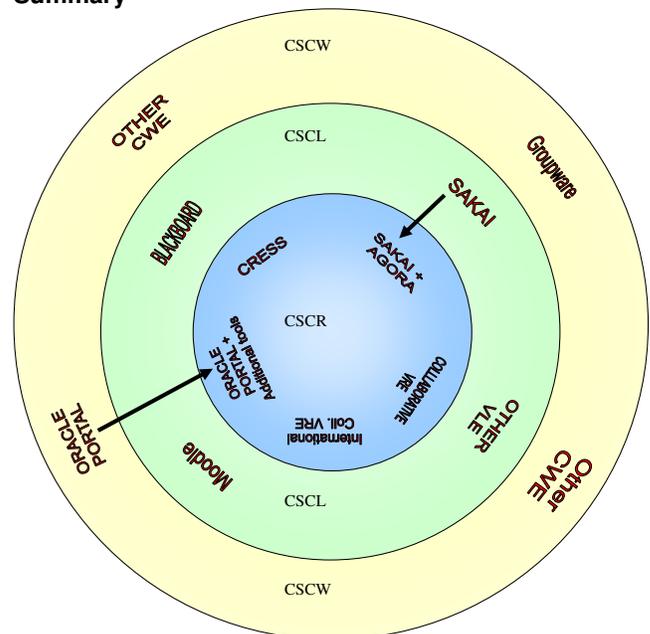

**Figure 1 Domain Diagram**

The purpose of this paper has been to find as closely matching set of CSCR tools within an existing portal framework as possible. An analysis of 10 portal frameworks has resulted in establishing

Sakai/Agora as the most applicable framework with only four tools missing from the package.

Figure 1 shows the relationship of the CSCW, CSCL and CSCR domains and the positioning of the various collaborative learning and research environments within those domains. It can be seen that some environments (e.g. Oracle portal) which have been designed for the CSCW domain can be useful within the richer CSCR domain provided that additional tools are developed. In particular attention is drawn to the Sakai Portal Framework which is suitable for use within the CSCL domain but with the addition of the Agora toolset and other portlets can be made suitable for the CSCR domain.

### Next stages

Future work will involve the building of a CRESS environment which will be based upon full usability analysis. Stage two will involve prototyping, initially in storyboard form, which will be submitted to potential users for initial usability feedback. A prototype will be produced from this and handed over to developers for the construction of the user interface package. This will lead onto usability testing to determine the adequacy of the user interface concepts. Once the basic framework has been established specific plug in modules may be incorporated for specific needs by specific groups. Lindgaard *et al* (2006) original methodology called for three iterations of design, prototype and usability test. However they were not able to maintain this in practice. It is envisaged that at least two or three iterations would be required to provide a stable and usable CRESS environment.

**VLE, VRE and Portal Frameworks**

http://agora.lancaster.ac.uk
http://www.blackboard.com
http://www.elgg.org
http://www.moodle.org
http://www.oracle.com
http://www.sakai.org
https://ugforge.ecs.soton.ac.uk
http://www.liferay.com
http://www.jboss.org
http://www.gridsphere.org
http://www.jporta.sourceforge.net
http://www.dotnetnuke.com
http://www.lightportal.org
http://www.cms.matrix.org

**Appendix**

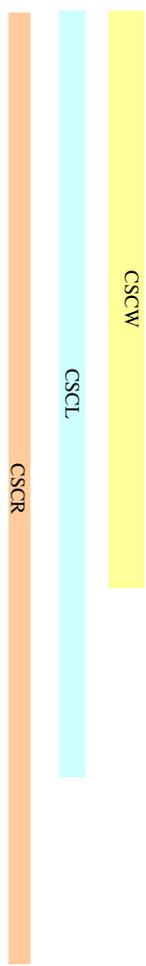

| CATEGORY SPACES | TOOLS | Argles SIM20 | Bachler "Buddy Space" | Baker et al "Grove Space" | Berger | Dalziel "LAMS" | Harper et al "ETR" | Hosoya et al "HyClass" | Kligyte et al "Fle3" | Miao et al "IMS-LD" | Pekkola "VIVA" | Walters et al "M-Grid" | Liccardi CAWS: A co-authoring Wiki | Sim et al Web/Grid Services VRE | Blackboard | Moodle | Elgg | total |
|---|---|---|---|---|---|---|---|---|---|---|---|---|---|---|---|---|---|---|
| Administration Space (including security tools) | Login | x | x | x | x | x | x | x | x | x | x | x | x | x | x | x | x | 16 |
| | Access/authorisation Tools | x | x | x | x | x | x | x | x | x | x | x | x | x | x | x | x | 16 |
| | Recording /Replay Facility | | x | | | | | | | | | | | | | | | 1 |
| | Instant Messaging Recording | | x | | | | | | | | | | | | | x | | 2 |
| | Assistive Agent | | | | | | x | | | | | | | | | | | 1 |
| | Help Pane | x | | x | | | | x | | | | | | | x | x | x | 6 |
| | Information Link Map | | x | | | | | x | x | | | | | | x | | | 4 |
| | Scenario/Control flow Tools | | | | | | | | | x | | | | | | | | 1 |
| Communication space (including Identification space) | Text/ Chat | x | | x | | x | x | | x | | | | | | x | | x | 8 |
| | Audio/Voice | | | x | | x | x | | x | | | | | | | | | 4 |
| | Still Picture | | | x | | | x | | | | | | | | x | x | x | 5 |
| | Video | | | x | | x | x | | | | | | | | | | | 3 |
| | Instant Messaging | | | x | | | | | x | | | | | | | x | | 3 |
| | Forum | | | | x | | | x | | | | | x | x | x | x | x | 7 |
| | Message Board/News | | | | | x | | | x | | | | x | | x | x | x | 6 |
| | Avatar (Representations) | | | | | | | x | | | | | | | | | | 1 |
| | Presence Indicator/Information | x | x | | | | | | x | | x | | | | | x | | 5 |
| | Location Identifier | | x | | | x | | | x | | | | | | | | | 3 |
| | Focus Indication | | x | | | x | | | x | | | | | | | | | 3 |
| | Participant Data | | x | | | | | | | | | | x | x | | x | x | 5 |
| Scheduling space | Scheduling Tool | x | | x | | x | | x | | x | | | | | x | x | | 7 |
| | Task Setting | | x | | | | | | | x | | | | | x | x | | 4 |
| | Task Monitoring | | x | x | | | | | x | | | | | x | x | | | 5 |
| Shared working space | Whiteboard | | x | | | | x | x | x | | | | | | x | | | 5 |
| | Collaborative Working Window | x | | x | | | x | x | | | | | x | | | | x | 6 |
| | 3D Environment | | | | | | | x | x | | | | | | | | | 2 |
| Product Space | Output Window | x | | | | | | | | | | | | x | | | | 2 |
| | Simulations | x | | | | | | | | | | | | x | x | | | 3 |
| Reflection Space | Reflective Journal/Private | | | | | x | | | | | | | x | | | x | x | 4 |
| Social Interaction Space | Community Creation | | | | | | | | | | | | x | | | | x | 2 |
| | Tags (marking Content) | | | | | | | | | | | | x | | | | x | 2 |
| | Friend (file sharing) | | | | | | | | | | | | x | | | | x | 2 |
| | Blog (Public + Private) | | | | | | | | | | | | x | x | | x | x | 4 |
| | RSS feed to centralize data | | | | | | | | | | | | | x | | | x | 2 |
| Assessment / Feedback Space | Assessment | | | | | | | | | | | | | | x | x | | 2 |
| | Feedback | | | | | | | | | | | | | | x | x | x | 3 |
| Supervisor Space | Private area for tutors | | | | | | | | | | | | | | x | x | x | 3 |
| Knowledge Space | Contribution Database | | | | x | | | x | x | | | | x | x | | | | 5 |
| | Academic database | | | | | | | | | | | | | x | | | | 1 |
| | Depository | | | | | x | | x | | x | | | x | x | x | x | x | 8 |
| | PowerPoint Slides | | x | | | | x | | | | | | | | x | x | | 4 |
| Privacy Space | Private Space | | | | | | | | | | | | | | x | x | x | 4 |
| Public Space | Public information space | | | | | | | | | | | | | | x | x | x | 3 |
| Negotiation Space | Peer Review assistance | | | | | | | | | | | | x | x | x | | | 3 |
| Publication Space | Schemas/Templates | | | | | | | | | | | | | | x | | x | 2 |
| | Publishing assistance | | | | | | | | | | | | | x | x | x | x | 4 |

**Table 3: Analysis of thirteen collaborative laboratories and three e-learning environments**

| CATEGORY SPACES | TOOLS | Ugforge | Liferay | JBoss | Gridsphere | Elgg | J Porta | DotNetNuke | Oracle | Light Portal | Sakai with Agora | CRESS Required | CRESS Not Required |
|---|---|---|---|---|---|---|---|---|---|---|---|---|---|
| | Count | 11 | 8 | 6 | 13 | 20 | 6 | 13 | 15 | 13 | 39 | 37 | |
| Administration Space (including security tools) | Login | X | X | X | X | X | X | X | X | | X | X | |
| | Access/authorisation Tools | | | | | | | | | | X | X | |
| | Recording /Replay Facility | | | | | | | | | | X | X | |
| | Instant Messaging Recording | | | | | | | | | | | | |
| | Assistive Agent | | | | | | | | | | | | X |
| | Help Pane | | X | X | | X | X | | | | X | X | |
| | Information Link Map | | | | | | | | | | | | X |
| | Scenario/Control flow Tools | | | | | | | | | | | | X |
| Communication space (including Identification space) | Text/ Chat | | | | X | | | | | X | X | X | |
| | Audio/Voice | | | | | | | | | | X | X | |
| | Still Picture | | X | | | X | | | | X | X | X | X |
| | Video | | | | | | | | X | | X | X | X |
| | Instant Messaging | | | | | | | | X | | X | | |
| | Forum | X | | | X | X | | X | X | X | X | | |
| | Message Board/News | X | | | | X | X | X | | X | X | X | |
| | Avatar (Representations) | | | | | | | | | | | | X |
| | Presence Indicator/Information | | | | | | | | | | X | X | |
| | Location Identifier | | | | | | | | | | | | X |
| | Focus Indication | | | | | | | | | | X | X | |
| | Participant Data | | X | | X | X | | X | | | X | X | |
| Scheduling space | Scheduling Tool (calendar) | X | X | | | | X | X | X | | X | X | |
| | Task Setting | X | | | | X | | | X | X | X | X | |
| | Task Monitoring | X | | | | X | | | X | X | X | X | |
| Shared working space | Whiteboard | | | | | | | | | | X | X | |
| | Collaborative Working Window /wiki | X | X | | | X | | X | X | | X | X | |
| | 3D Environment | | | | | | | | | | | | X |
| Product Space | Output Window | | | | | | | | | | X | | |
| | Simulations | | | | | | | | | | | | X |
| Reflection Space | Reflective Journal/Private | | | | | | | | | | | X | |
| Social Interaction Space | Community Creation | | X | X | X | | | | | | X | X | |
| | Tags (marking Content) | | X | | | X | | | | | | X | |
| | Friend (file sharing) | | | | | | | | | | | X | |
| | Blog (Public + Private) | | | | X | X | | X | | | X | X | |
| | RSS feed to centralized data | | | | X | X | X | X | | X | X | X | |
| Assessment / Feedback Space | Assessment | | | | | | | | | | X | X | |
| | Feedback | | | | | | | X | | X | X | X | |
| Supervisor Space | Private area for tutors | | | | | | | | | | X | X | |
| Knowledge Space | Contribution Database | | | | | | | | X | | | X | |
| | Academic database | | | | | | | | X | | | X | |
| | Depository (shared files) | X | | X | | X | | X | X | X | X | X | |
| | PowerPoint Slides | | | | | | | | | | X | X | |
| Privacy Space | Private Space | | | | | | | | | | X | X | |
| Public Space | Public information space | | | | | | | | | | X | X | |
| Negotiation | Peer Review assistance | | | | | | | | | | | X | |
| Publication Space | Schemas/Templates (doc archive) | X | | | | X | | | | | X | X | |
| | Publishing assistance | | | | | | | | X | | | X | |
| OTHER | Layout customization | | | | X | X | X | X | | X | | X | |
| | email | | | | X | X | | | | | | X | |
| | Search | | | | | | | X | | | | | |
| | Banner | | | | | | | X | | | | | |
| | Still image slides (Gallery) | | | | | X | X | X | | | | | |
| | Lists/Links | | | | | X | X | X | X | X | | | |
| | Mobile Device Support (including Pod) | | X | | | X | | | | | | | |
| | Themes | | X | | | X | | | | | | | |
| | User surveys | X | | | | | | | | X | | | |
| | Feature request tracking | X | | | | | | | | | | | |
| | Bug tracking | X | | | | | | | | | | | |
| | External Websites | | | | | | | | X | X | X | | |
| | Manage Groups | | | | | | | | | | X | | |
| | Tests and Quiz | | | | | | | | | | X | | |
| | Web content | | | | | | | | | | X | | |
| | Worksite set-up | | | | | | | | | | X | | |
| | Syllabus | | | | | | | | | | X | | |
| | Movie casting | | | | | | | | | | X | | |

**Table 4 Portal Framework Analysis**

| CSCR Categories | ALREADY AVAILABLE in Sakai/Agora | Corresponding Tool in Sakai / Agora | NOT AVAILABLE In Sakai/Agora | Required by CRESS |
|---|---|---|---|---|
| Administration Space (including security tools) | Login | Sakai: Permissions and Roles | | X |
| | Access/authorisation Tools | Sakai: Permissions and Roles | | X |
| | Recording /Replay Facility | Agora: Session recording | | X |
| | Help Pane | Sakai: Help tool | | x |
| Communication tools (including Identification tools) | Text/ Chat | Sakai: Chat room; Agora: Chat | | x |
| | Audio/Voice | Agora: Video conferencing | | x |
| | Still Picture | Sakai: Profile | | x |
| | Video | Agora: Video conferencing | | x |
| | Forum | Sakai: Discussion tool | | x |
| | Message Board/News | Sakai: Announcement tool | | x |
| | Presence Indicator/Information | Agora: Video conferencing | | x |
| | Focus Indication | Agora: Video conferencing | | x |
| | Participant Data | Sakai: Profile | | x |
| Scheduling | Scheduling Tool (calendar) | Sakai: Schedule tool | | x |
| | Task Setting | Sakai: My Workspace | | x |
| | Task Monitoring | Sakai: My Workspace | | x |
| Shared | Whiteboard | Agora: Shared Desktop | | x |
| | Collaborative Working Window (wiki) | Sakai: Wiki tool | | x |
| Product | Output Window | Agora: Shared Desktop | | x |
| Reflection | Reflective Journal/Private | Sakai: My Workspace | | x |
| Social Interaction | Community Creation | Sakai Membership tool | | x |
| | | | Tags (marking Content) | x |
| | Friend (file sharing) | Sakai: Resources tool | | x |
| | Blog (Public + Private) | Sakai: Wiki tool | | x |
| | RSS feed to centralized data | Sakai: News tool | | x |
| Assessment / Feedback | Assessment | Sakai: Post'em | | x |
| | Feedback | Sakai: Post'em | | x |
| Supervisor | Private area for tutors | Sakai: Discussion tool | | x |
| Knowledge | | | Contribution Database | x |
| | Academic database (Google scholar etc.)tool | Sakai: Web content | | x |
| | Depository (shared files) | Sakai: Drop Box tool | | x |
| | PowerPoint Slides | Sakai: Drop Box tool | | x |
| Privacy | Private Space | Sakai: My workspace | | x |
| Public | Public information space | Sakai: Site Info tool | | x |
| Negotiation | | | Peer Review assistance | x |
| Publication | Schemas/Templates (doc archive) | Sakai: Resources tool | | x |
| | | | Publishing assistance | x |
| Additional Features available in Sakai | Layout customization<br>email<br>Search<br>Banner<br>Still image slides (Gallery)<br>Lists/Links<br>Mobile Device Support (including Pod) casting etc)<br>Themes<br>User surveys<br>Feature request tracking<br>Bug tracking<br>External Websites<br>Manage Groups<br>Tests and Quiz<br>Web content<br>Worksite set-up<br>Syllabus<br>Movie casting | | | |

**Table 5 Sakai/Agora Tool Gap Analysis with CRESS requirements**